\documentclass[12pt]{iopart}
\usepackage{iopams}  

\usepackage[harvard,dcucite]{harvard}

\newcommand{\C}{\mathbb{C}}

\newcommand{\hache}{\mathbb{H}}

\newcommand{\Z}{\mathbb{Z}}

\newcommand{\intv}[2]{[\![#1,#2]\!]}

\newcommand{\drc}[1]{\left|#1\right\rangle}

\newenvironment{tabcen}[1]{\begin{center}\begin{tabular}{#1}}{\end{tabular}\end{center}}
 
\newcommand{\proto}[3]{\begin{tabcen}{p{.4\textwidth}cp{.4\textwidth}}
\hspace{1.8cm}{#1} & & \hspace{1.8cm}{#2} \\ \hline
#3
\hline\end{tabcen}}

\begin{document}

\title[Communication protocols based on entanglement swapping]{Quantum communication protocols based on entanglement swapping}

\author{Guillermo Morales-Luna}

\address{Computer Science Department, CINVESTAV-IPN, Mexico City, Mexico}
\ead{gmorales@cs.cinvestav.mx}
\vspace{10pt}
\begin{indented}
\item[]January 2015
\end{indented}

\begin{abstract}
We recall several cryptographic protocols based on entanglement alone and also on entanglement swapping. We make an exposition in terms of the geometrical aspects of the involved Hilbert spaces, and we concentrate on the formal nature of the used transformations.
\end{abstract}

\section{Introduction}

Entanglement has been widely exploited in the design of protocols and procedures for communication, cryptography and computation within quantum contexts. Quantum codes guarantee that information has been transmitted without any alteration. Cryptographic protocols aid for key agreement of for secure communication, namely unconditionally secure information exchange. Entanglement has been used to implement and to speed-up paradigmatic quantum algorithms~\cite{Gisin02,Lanyon07}.

Entanglement swapping may entangle two quantum systems without direct interaction among them, and this fact is exploited within several  quantum cryptography schemes. 

Here, we recall several well known cryptographic protocols using entanglement, alone, and entanglement swapping: the {\em Quantum Secure Direct Communication Protocol} (see table~\ref{tb:QSDC} below) communicates securely bit strings with an even length, the {\em Quantum Bidirectional Communication Protocol} (see table~\ref{tb:QBC1} below) is a generalization of the above protocol in which the communicating parts exchange simultaneously messages of even bit length, the {\em Quantum Multidirectional Communication Protocol} (see table~\ref{tb:QBC2} below) allows the message exchange among three parts using the maximally entangled GHZ states, at table~\ref{tb:QCBCk} we recall a three parties protocol in which two correspondents communicate securely just after the authorisation of a third party (who does not catch the message exchange), and finally, the {\em Key Agreement Protocol Using Entanglement Swapping} is sketched at table~\ref{tb:kapues}  in order to illustrate the use of entanglement swapping in cryptographic protocols.

We emphasise the algebraic aspect of the Hilbert space nature of all the involved protocols. We establish a correspondence among unitary transforms, obtained as tensor products of Pauli maps, and permutations of basic vectors in the Hilbert spaces. These correspondence are summarised at tables~\ref{tb:A2} and~\ref{tb:A3}. Also, explicit expressions of the Bell basis, in terms of the Hadamard basis are given.

For any two integer numbers $i,j\in\Z^+$, with $i\leq j$, let us write 
$$\intv{i}{j} = \{i,i+1,\ldots,j-1,j\}.$$

\section{Qubits and Pauli transforms}\label{sc.swit}

Let $\hache_1=\C^2$ be the two-dimensional complex Hilbert space. The unit sphere of $\hache_1$ is the set of {\em qubits}. The {\em canonical basis} consists of the vectors ${\bf e}_0 = [1\ 0]^T$ and ${\bf e}_1 = [0\ 1]^T$. Usually, it is written $\drc{0}={\bf e}_0$ and $\drc{1}={\bf e}_1$. 
Let ${\bf h}_0 = \frac{1}{\sqrt{2}}(\drc{0}+\drc{1})$ and ${\bf h}_1 = \frac{1}{\sqrt{2}}(\drc{0}-\drc{1})$ be the vectors forming the {\em Hadamard basis} at $\hache_1$. 

Let us consider the Pauli operators
\begin{eqnarray}
\sigma_0 = \left(\begin{array}{rr}
 1 &  0 \\
 0 &  1 
\end{array}\right) &,&\ \ 
\sigma_x = \left(\begin{array}{rr}
 0 &  1 \\
 1 &  0 
\end{array}\right) \vspace{2ex} \nonumber  \\
\sigma_y = \left(\begin{array}{rr}
 0 & -i \\
 i &  0 
\end{array}\right) &,&\ \ 
\sigma_z = \left(\begin{array}{rr}
 1 &  0 \\
 0 & -1 
\end{array}\right)
 \label{eq.pauli}
\end{eqnarray}
and let us number them as $[\sigma_0\ \ \sigma_1\ \ \sigma_2\ \ \sigma_3] = [\sigma_0\ \ \sigma_x\ \ \sigma_y\ \ \sigma_z]$.
The action of these operators over the vectors at the canonical and the Hadamard basis is sketched at table~\ref{tb:actpau}. At each entry is located the value of $\sigma_i$ at the vector labeling the corresponding column.
\begin{table}
\caption{{\em Action of the Pauli operators on basis vectors} \label{tb:actpau}}
$$\begin{array}{c||r|r|r|r||} 
 & \drc{0} & \drc{1} & {\bf h}_0 & {\bf h}_1 \\ \hline \hline
\sigma_0 & \drc{0} & \drc{1} & {\bf h}_0 & {\bf h}_1 \\ \hline 
\sigma_1 & \drc{1} & \drc{0} & {\bf h}_0 & -{\bf h}_1 \\ \hline 
\sigma_2 & -i\,\drc{1} & i\,\drc{0} & i\,{\bf h}_1 & -i\,{\bf h}_0 \\ \hline 
\sigma_3 & \drc{0} & -\drc{1} & {\bf h}_1 & {\bf h}_0 \\ \hline \hline 
\end{array}$$
\end{table}
We see that, up to a factor which is an unitary complex number, the canonical basis remains fixed by $\sigma_0$ and $\sigma_3$ and is switched by $\sigma_1$ and $\sigma_2$ while the Hadamard basis remains fixed by $\sigma_0$ and $\sigma_1$ and is switched by $\sigma_2$ and $\sigma_3$. Thus, the operator $\sigma_2$ is switching both basis. 

Let $\hache_2=\hache_1\otimes\hache_1$ be the Hilbert space containing the 2-quregisters. Any 2-quregister ${\bf x}$ has two components ${\bf x}_0$ and ${\bf x}_1$, each at the factor space $\hache_1$, they are qubits. 
Let for $i,j\in\intv{0}{1}$, 
$${\bf b}_{ij} = \frac{1}{\sqrt{2}}\left(\drc{0i}+(-1)^j\drc{1\overline{i}}\right),$$
here the overline denotes complement modulus 2. Then $B=\left({\bf b}_{ij}\right)_{i,j\in\intv{0}{1}}$ is the Bell basis of $\hache_2$ and consists of four maximally entangled states. 
In terms of the Hadamard basis, the Bell vectors are expressed as follows:
\begin{eqnarray*}
{\bf b}_{00} &=& \frac{1}{\sqrt{2}}({\bf h}_0\otimes{\bf h}_0+{\bf h}_1\otimes{\bf h}_1) \\
{\bf b}_{01} &=& \frac{1}{\sqrt{2}}({\bf h}_1\otimes{\bf h}_0+{\bf h}_0\otimes{\bf h}_1) \\
{\bf b}_{10} &=& \frac{1}{\sqrt{2}}({\bf h}_0\otimes{\bf h}_0-{\bf h}_1\otimes{\bf h}_1) \\
{\bf b}_{11} &=& \frac{1}{\sqrt{2}}({\bf h}_1\otimes{\bf h}_0-{\bf h}_0\otimes{\bf h}_1) 
\end{eqnarray*}

Any sequence $C=\left({\bf c}_k\right)_{k\geq 0}$ whose terms are elements of $B$ determines two sequences of qubits $C_0=\left({\bf c}_{k0}\right)_{k\geq 0}$ and $C_1=\left({\bf c}_{k1}\right)_{k\geq 0}$. 

Through the radix expression of an index in base 2, we may number the Bell basis as $B=\left({\bf b}_k\right)_{k\in\intv{0}{2^2-1}}$. The tensor products $\sigma_{ij}=\sigma_i\otimes\sigma_j$ determine bijections $\alpha_{ij}:B\to B$, such that
\begin{equation}
\forall (i,j)\in\intv{0}{3}^2\ \forall k\in\intv{0}{2^2-1}:\ \ \sigma_{ij}({\bf b}_k) \in{\cal L}({\bf b}_{\alpha_{ij}(k)}). \label{eq.bij}
\end{equation}
Let $A_2:\sigma_{ij}\mapsto\alpha_{ij}$ be the map that associates to each  tensor product $\sigma_{ij}$ the corresponding permutation that it defines at the Bell basis. The image of $A_2$ consists of just $4=2^2$  permutations $\left(\beta_{\ell}\right)_{\ell\in\intv{0}{2^3-1}}$, and each permutation is defined by 4  tensor products $\sigma_{ij}$ as summarized at the table~\ref{tb:A2}: the first column displays the index $\ell$, the second column the permutation $\beta_{\ell}$ and the third column the list of tensor product maps $\sigma_{ij}$ producing $\beta_{\ell}$ under the map $A_2$.
\begin{table}
\caption{{\em Correspondence on permutations of the Bell basis and tensor products of Pauli operators} \label{tb:A2}}
$$\begin{array}{||c|c|c||} \hline \hline
\ell & \beta_{\ell} & A_2^{-1}(\beta_{\ell}) \\ \hline \hline
0 & [0\ \ 1\ \ 2\ \ 3] & \{\sigma_{00},\sigma_{01},\sigma_{02},\sigma_{03}\} \\ \hline
1 & [2\ \ 3\ \ 0\ \ 1] & \{\sigma_{01},\sigma_{00},\sigma_{03},\sigma_{02}\} \\ \hline
2 & [3\ \ 2\ \ 1\ \ 0] & \{\sigma_{02},\sigma_{03},\sigma_{00},\sigma_{01}\} \\ \hline
3 & [1\ \ 0\ \ 3\ \ 2] & \{\sigma_{03},\sigma_{02},\sigma_{01},\sigma_{00}\} \\ \hline \hline
\end{array}$$
\end{table}

The table~\ref{tb:A2} can in turn be summarized as
\begin{eqnarray}
\forall i\in\intv{0}{3}:\ \ A_2(\sigma_{i\beta_0(i)})=\beta_0 &,&\ \  A_2(\sigma_{i\beta_3(i)})=\beta_1 , \nonumber \\
\hspace{2.5cm}A_2(\sigma_{i\beta_1(i)})=\beta_2 &,&\ \  A_2(\sigma_{i\beta_2(i)})=\beta_3.
\end{eqnarray}
By looking at relation~(\ref{eq.bij}), we see that if $i$ and $k$ remain fixed, then the index $j$ can be encoded by the value $\alpha_{ij}(k)$. This property can be exploited for secure communication purposes.

In table~\ref{tb:QSDC}  a  {\em Quantum Secure Direct Communication Protocol}~\cite{Deng03} is sketched. The purpose of this protocol is to communicate securely a word in $\intv{0}{3}^*$. 
Alice should communicate a message $\left[\mu_{\kappa}\right]_{\kappa=0}^{m-1}\in\intv{0}{3}^m$.
Alice and Bob fix a Pauli transform $\sigma_i$, a Bell quregister ${\bf b}_k\in B$, an integer $n>m$ and an index set $J\in\intv{0}{n-1}^{(m)}$. They share initially a constant sequence $C=\left({\bf c}_{\kappa}\right)_{{\kappa}= 0}^{n-1}$ whose terms coincide with ${\bf b}_k\in B$.

\begin{table}
\caption{{\em Quantum Secure Direct Communication Protocol} \label{tb:QSDC}}
\centering\fbox{\begin{minipage}{.95\textwidth}
\proto{Alice}{Bob}{
pads the message $\left[\mu_{\kappa}\right]_{\kappa=0}^{m-1}$ into a sequence $\left[j_{\kappa}\right]_{\kappa=0}^{n-1}$ by inserting the message into the positions at $J$  &  &   \\
codifies the message $\left[j_{\kappa}\right]_{\kappa=0}^{n-1}$ by calculating $D_1=\left(\sigma_{j_{\kappa}}({\bf c}_{\kappa 1})\right)_{{\kappa}= 0}^{n-1}$  &  &   \\
sends $D_1$ through a quantum channel & $\stackrel{D_1}{\longrightarrow}$ & receives $D_1$  \\
 &  & calculates $D_0=\left(\sigma_{i}({\bf c}_{\kappa 0})\right)_{{\kappa}= 0}^{n-1}$, which actually completes the calculation of $D=\left((\sigma_i\otimes\sigma_{j_{\kappa}})({\bf c}_{\kappa})\right)_{{\kappa}= 0}^{n-1}$ \\
 &  & calculates $E$ by measuring each term at $D$ with respect to the Bell basis  \\
 &  & recovers the sequence $\left(\alpha_{ij_{\kappa}}(k)\right)_{{\kappa}= 0}^{n-1}$, consequently the padded sequence $\left[j_{\kappa}\right]_{\kappa=0}^{n-1}$, and the original message $\left[\mu_{\kappa}\right]_{\kappa=0}^{m-1}$ \\
} 
\end{minipage}}
\end{table}

Using sequences of entangled quregisters it is also possible to build bidirectional communication protocols. In table~\ref{tb:QBC1}  a {\em Quantum Bidirectional Communication Protocol} is sketched~\cite{Gao08}. The purpose of this protocol is to communicate securely two words in $\intv{0}{3}^*$, one going from Alice to Bob and the other in the opposite direction. Alice and Bob should interchange messages in $\intv{0}{1}^{2n}$, and they share initially a constant sequence $C=\left({\bf c}_{\kappa}\right)_{{\kappa}= 0}^{n-1}$.
\begin{table}
\caption{{\em Quantum Bidirectional Communication Protocol} \label{tb:QBC1}}
\centering\fbox{\begin{minipage}{.95\textwidth}
\proto{Alice}{Bob}{
by taking pairs of contiguous bits, she writes her message as a word $\left(i_{\kappa}\right)_{{\kappa}= 0}^{n-1}\in\intv{0}{3}^n$  &  & by taking pairs of contiguous bits, he writes his message as a word $\left(j_{\kappa}\right)_{{\kappa}= 0}^{n-1}\in\intv{0}{3}^n$  \\
codifies her message by calculating $D_0=\left(\sigma_{i_{\kappa}}({\bf c}_{\kappa 0})\right)_{{\kappa}= 0}^{n-1}$ &  & codifies his message by calculating $D_1=\left(\sigma_{j_{\kappa}}({\bf c}_{\kappa 1})\right)_{{\kappa}= 0}^{n-1}$  \\
sends $D_0$ through a quantum channel & $\stackrel{D_0}{\longrightarrow}$ & receives $D_0$  \\
receives $D_1$ & $\stackrel{D_1}{\longleftarrow}$ & sends $D_1$  through a quantum channel \\
for each $\kappa\in\intv{0}{n-1}$ she measures the entangled quregister ${\bf d}_{\kappa}$ with respect to the Bell basis &  & for each $\kappa\in\intv{0}{n-1}$ he measures the entangled quregister ${\bf d}_{\kappa}$ with respect to the Bell basis  \\
since she knows $i_{\kappa}$, using~(\ref{eq.bij}), she recovers $j_{\kappa}$ &  & since he knows $j_{\kappa}$, using~(\ref{eq.bij}), he recovers $i_{\kappa}$  \\
}
\end{minipage}}
\end{table}

\section{Entanglement swapping}

{\em Entanglement swapping} is a phenomenon which allows to put two particles into entangled states although these particles have not been close at any time. Departing from two pairs of entangled particles, a particle is chosen from each pair, then the joint pair of selected particles is measured with respect to the Bell basis, resulting in an entangled state. As a consequence, the pair consisting of the two partner particles is also entangled. This last pair is the result of the entanglement swapping beginning from the original two pairs.

In two 2-quregisters there are involved 4 qubits, let us identify them with the four indexes in the integer interval $\intv{0}{3}$. Let us write $\drc{\varepsilon}_{\mu}$, for $\varepsilon\in\{0,1\}$, $\mu\in\intv{0}{3}$, to denote the state of the $\mu$-th qubit. For any two different indexes $\mu,\nu\in\intv{0}{3}$, let the respective Bell basis of the Hilbert space $\hache_2$ be
$$\left({\bf b}^{(\mu\nu)}_{ij} = \frac{1}{\sqrt{2}}\left(\drc{0i}_{\mu\nu}+(-1)^j\drc{1\overline{i}}_{\mu\nu}\right)\right)_{i.j\in\intv{0}{1}}.$$
Let us assume that the 2-quregister consisting of the qubits 0 and 1 is entagled as well as the pair of qubits 2 and 3. Then a basis of the space $\hache_4$ is
\begin{equation}
B_{(01)(23)} = \left({\bf z}_{i_0j_0i_1j_1}={\bf b}^{(01)}_{i_0j_0} \otimes {\bf b}^{(23)}_{i_1j_1}
\right)_{i_0,j_0,i_1,j_1\in\intv{0}{1}}, \label{eq.es01}
\end{equation}
where $\forall i_0,j_0,i_1,j_1\in\intv{0}{1}$:
\begin{eqnarray}
{\bf z}_{i_0j_0i_1j_1} &=&  \frac{1}{2}
\left(\hspace{3em}\drc{0i_00i_1}_{0123}+(-1)^{j_1}\drc{0i_01\overline{i_1}}_{0123}+\right. \nonumber \\
 & & \hspace{1em} \left.(-1)^{j_0}\drc{1\overline{i_0}0i_1}_{0123}+(-1)^{j_0+j_1}\drc{1\overline{i_0}1\overline{i_1}}_{0123}\right). \label{eq.es02}
\end{eqnarray}
By rearranging the pairs and considering the pairs $(0,2)$ and $(1,3)$, we have that a second basis of $\hache_4$ is
\begin{equation}
B_{(02)(13)} = \left({\bf b}^{(02)}_{i_0j_0} \otimes {\bf b}^{(13)}_{i_1j_1}
\right)_{i_0,j_0,i_1,j_1\in\intv{0}{1}}, \label{eq.es03}
\end{equation}
where $\forall i_0,j_0,i_1,j_1\in\intv{0}{1}$:
\begin{eqnarray*}
{\bf b}^{(02)}_{i_0j_0} \otimes {\bf b}^{(13)}_{i_1j_1} &=& \frac{1}{2}
\left(\hspace{3em}\drc{0i_00i_1}_{0213}+(-1)^{j_1}\drc{0i_01\overline{i_1}}_{0213}+\right. \\
 & & \hspace{1em}\left.(-1)^{j_0}\drc{1\overline{i_0}0i_1}_{0213}+(-1)^{j_0+j_1}\drc{1\overline{i_0}1\overline{i_1}}_{0213}\right).
\end{eqnarray*}
By swapping the middle qubits, the following 4-quregisters result: $\forall i_0,j_0,i_1,j_1\in\intv{0}{1}$,
\begin{eqnarray}
{\bf y}_{i_0j_0i_1j_1} &=& \frac{1}{2}
\left(\hspace{3em}\drc{00i_0i_1}_{0123}+(-1)^{j_1}\drc{01i_0\overline{i_1}}_{0123}+\right.  \nonumber \\
 & & \hspace{1em}\left.(-1)^{j_0}\drc{10\overline{i_0}i_1}_{0123}+(-1)^{j_0+j_1}\drc{11\overline{i_0}\overline{i_1}}_{0123}\right). \label{eq.es04}
\end{eqnarray}
Each 2-quregister ${\bf z}_{i_0j_0i_1j_1}$ given by relation~(\ref{eq.es02}) can be expressed in terms of the 2-quregisters ${\bf y}_{i_0j_0i_1j_1}$ given by relation~(\ref{eq.es04}), namely:
\begin{eqnarray}
{\bf z}_{i_0j_0i_1j_1} &=& \frac{1}{2}
\left(\hspace{3em}{\bf y}_{i_0j_0i_1j_1}+(-1)^{j_1}{\bf y}_{i_0\overline{j_0}i_1\overline{j_1}}+ \right. \nonumber \\
 & &  \hspace{1em}\left. (-1)^{j_0}{\bf y}_{\overline{i_0}j_0\overline{i_1}j_1}+(-1)^{j_0+j_1}{\bf y}_{\overline{i_0}\overline{j_0}\overline{i_1}\overline{j_1}}\right), \label{eq.es05}
\end{eqnarray}
and this relation is symmetric:
\begin{eqnarray}
{\bf y}_{i_0j_0i_1j_1}  &=&  \frac{1}{2}
\left(\hspace{3em}{\bf z}_{i_0j_0i_1j_1}+(-1)^{j_1}{\bf z}_{i_0\overline{j_0}i_1\overline{j_1}}+ \right.  \nonumber \\
 & & \hspace{1em}\left.(-1)^{j_0}{\bf z}_{\overline{i_0}j_0\overline{i_1}j_1}+(-1)^{j_0+j_1}{\bf z}_{\overline{i_0}\overline{j_0}\overline{i_1}\overline{j_1}}\right), \label{eq.es06}
\end{eqnarray}
In this way, the entanglement of the 4-registers ${\bf z}$ is reflected by the entanglement of the 4-registers ${\bf y}$, in other words, the entanglement of the pairs $(0,1)$ and $(2,3)$ is swapped into the entanglement of the pairs $(0,2)$ and $(1,3)$, and conversely.

\section{Three-entanglement}

Let us consider multi-party bidirectional protocols. In particular, we will illustrate these procedures with three communicating parties. 
A proper protocol considers maximally entangled 3-quregisters, members of $\hache_3=\hache_1\otimes\hache_2$.  Any 3-quregister ${\bf x}$ has three components ${\bf x}_0$, ${\bf x}_1$ and ${\bf x}_2$, each at the factor space $\hache_1$, they are qubits. For $\varepsilon_1,\varepsilon_2,\varepsilon_3\in\intv{0}{1}$ let
$${\bf b}_{\varepsilon_1\varepsilon_2\varepsilon_3} = \frac{1}{\sqrt{2}}\left(\drc{0\varepsilon_1\varepsilon_2} + (-1)^{\varepsilon_3}\drc{1\overline{\varepsilon_1}\overline{\varepsilon_2}}\right).$$
These vectors form a basis, $B_3$, analogous to the Bell basis in $\hache_3$, but they are called {\em Greensberger-Horne-Zeilinger} (GHZ) {\em states}. In terms of the Hadamard vectors, the GHZ states are expressed as shown at the table~\ref{tb:ghz}.
\begin{table}
\caption{The GHZ states in terms of the Hadamard basis of qubits. \label{tb:ghz}}
\centering\fbox{\begin{minipage}{.95\textwidth}
\begin{eqnarray*}
{\bf b}_{000} &=& \frac{1}{2}\left({\bf h}_0\otimes\left({\bf h}_0\otimes{\bf h}_0+{\bf h}_1\otimes{\bf h}_1\right)+
{\bf h}_1\otimes\left({\bf h}_0\otimes{\bf h}_1+{\bf h}_1\otimes{\bf h}_0\right)\right) \\
{\bf b}_{001} &=& \frac{1}{2}\left({\bf h}_1\otimes\left({\bf h}_0\otimes{\bf h}_0+{\bf h}_1\otimes{\bf h}_1\right)+
{\bf h}_0\otimes\left({\bf h}_0\otimes{\bf h}_1+{\bf h}_1\otimes{\bf h}_0\right)\right) \\
{\bf b}_{010} &=& \frac{1}{2}\left({\bf h}_0\otimes\left({\bf h}_0\otimes{\bf h}_0-{\bf h}_1\otimes{\bf h}_1\right)+
{\bf h}_1\otimes\left(-{\bf h}_0\otimes{\bf h}_1+{\bf h}_1\otimes{\bf h}_0\right)\right) \\
{\bf b}_{011} &=& \frac{1}{2}\left({\bf h}_1\otimes\left({\bf h}_0\otimes{\bf h}_0-{\bf h}_1\otimes{\bf h}_1\right)+
{\bf h}_0\otimes\left(-{\bf h}_0\otimes{\bf h}_1+{\bf h}_1\otimes{\bf h}_0\right)\right) \\
{\bf b}_{100} &=& \frac{1}{2}\left({\bf h}_0\otimes\left({\bf h}_0\otimes{\bf h}_0-{\bf h}_1\otimes{\bf h}_1\right)+
{\bf h}_1\otimes\left({\bf h}_0\otimes{\bf h}_1-{\bf h}_1\otimes{\bf h}_0\right)\right) \\
{\bf b}_{101} &=& \frac{1}{2}\left({\bf h}_1\otimes\left({\bf h}_0\otimes{\bf h}_0-{\bf h}_1\otimes{\bf h}_1\right)+
{\bf h}_0\otimes\left({\bf h}_0\otimes{\bf h}_1-{\bf h}_1\otimes{\bf h}_0\right)\right) \\
{\bf b}_{110} &=& \frac{1}{2}\left({\bf h}_0\otimes\left({\bf h}_0\otimes{\bf h}_0+{\bf h}_1\otimes{\bf h}_1\right)+
{\bf h}_1\otimes\left(-{\bf h}_0\otimes{\bf h}_1-{\bf h}_1\otimes{\bf h}_0\right)\right) \\
{\bf b}_{111} &=& \frac{1}{2}\left({\bf h}_1\otimes\left({\bf h}_0\otimes{\bf h}_0+{\bf h}_1\otimes{\bf h}_1\right)+
{\bf h}_0\otimes\left(-{\bf h}_0\otimes{\bf h}_1-{\bf h}_1\otimes{\bf h}_0\right)\right) 
\end{eqnarray*}
\end{minipage}}
\end{table}

Through the radix expression of an index in base 2, we may number the Bell basis as $B_3=\left({\bf b}_k\right)_{k\in\intv{0}{2^3-1}}$. The tensor products $\sigma_{ijk}=\sigma_i\otimes\sigma_j\otimes\sigma_k$ determine bijections $\alpha_{ijk}:B_3\to B_3$ in an analogous way as in~(\ref{eq.bij}):
\begin{equation}
\forall (i,j,k)\in\intv{0}{3}^3\ \forall \ell\in\intv{0}{2^3-1}:\ \ \sigma_{ijk}({\bf b}_{\ell}) \in{\cal L}({\bf b}_{\alpha_{ijk}(\ell)}). \label{eq.bij3}
\end{equation}
Let $A_3:\sigma_{ijk}\mapsto\alpha_{ijk}$ be the map that associates to each  tensor product $\sigma_{ijk}$ the corresponding permutation at the Bell basis. The image of $A_3$ consists of just $8=2^3$  permutations $\left(\beta_{\mu}\right)_{\mu\in\intv{0}{2^3-1}}$, and each permutation is defined by 8  tensor products $\sigma_{ijk}$ as summarized in table~\ref{tb:A3}.
\begin{table}
\caption{{\em Correspondence on permutations of the Bell basis and tensor products of Pauli operators} \label{tb:A3}}
$$\begin{array}{||c|c|c||} \hline \hline
\mu & \beta_{\mu} & A_3^{-1}(\beta_{\mu}) \\ \hline \hline
0 & [0\ \ 1\ \ 2\ \ 3\ \ 4\ \ 5\ \ 6\ \ 7] & \{\sigma_{000},\sigma_{033},\sigma_{111},\sigma_{122}, \sigma_{212},\sigma_{221},\sigma_{303},\sigma_{330}\} \\ \hline
1 & [2\ \ 3\ \ 0\ \ 1\ \ 6\ \ 7\ \ 4\ \ 5] & \{\sigma_{003},\sigma_{030},\sigma_{112},\sigma_{121}, \sigma_{211},\sigma_{222},\sigma_{300},\sigma_{333}\} \\ \hline
2 & [3\ \ 2\ \ 1\ \ 0\ \ 7\ \ 6\ \ 5\ \ 4] & \{\sigma_{001},\sigma_{032},\sigma_{110},\sigma_{123}, \sigma_{213},\sigma_{220},\sigma_{302},\sigma_{331}\} \\ \hline
3 & [1\ \ 0\ \ 3\ \ 2\ \ 5\ \ 4\ \ 7\ \ 6] & \{\sigma_{002},\sigma_{031},\sigma_{113},\sigma_{120}, \sigma_{210},\sigma_{223},\sigma_{301},\sigma_{332}\} \\ \hline
4 & [4\ \ 5\ \ 6\ \ 7\ \ 0\ \ 1\ \ 2\ \ 3] & \{\sigma_{010},\sigma_{023},\sigma_{101},\sigma_{132}, \sigma_{202},\sigma_{231},\sigma_{313},\sigma_{320}\} \\ \hline
5 & [6\ \ 7\ \ 4\ \ 5\ \ 2\ \ 3\ \ 0\ \ 1] & \{\sigma_{013},\sigma_{020},\sigma_{102},\sigma_{131}, \sigma_{201},\sigma_{232},\sigma_{310},\sigma_{323}\} \\ \hline
6 & [7\ \ 6\ \ 5\ \ 4\ \ 3\ \ 2\ \ 1\ \ 0] & \{\sigma_{011},\sigma_{022},\sigma_{100},\sigma_{133}, \sigma_{203},\sigma_{230},\sigma_{312},\sigma_{321}\} \\ \hline
7 & [5\ \ 4\ \ 7\ \ 6\ \ 1\ \ 0\ \ 3\ \ 2] & \{\sigma_{012},\sigma_{021},\sigma_{103},\sigma_{130}, \sigma_{200},\sigma_{233},\sigma_{311},\sigma_{322}\} \\ \hline
 \hline
\end{array}$$
\end{table}
As seen at the beginning of section~\ref{sc.swit}, the operator $\sigma_2$ switches the canonical and the Hadamard basis. Let us consider just operators of the form $\sigma_{ijk}$ where $i\in\intv{0}{3}$, $j,k\in\{0,2\}$.
Then, the restriction of table~\ref{tb:A3} to operators at the set $S=\{\sigma_{ijk}|\ i\in\intv{0}{3},j,k\in\{0,2\}\}$ is shown at table~\ref{tb:SA}.
\begin{table}
\caption{{\em Correspondence on permutations of the Bell basis and tensor products of Pauli operators at the set $S$, in alphabetical order according to the first element at $S\cap A_3^{-1}(\beta_{\mu})$. Observe that if one index $i,j,k$ is known, the other two can be deduced form this index and the $3$-quregister $\sigma_{ijk}({\bf b}_{\ell})$.} \label{tb:SA}}
$$\begin{array}{||c|c||} \hline \hline
\mu & S\cap A_3^{-1}(\beta_{\mu}) \\ \hline \hline
0 & \{\sigma_{000},\sigma_{122}\} \\ \hline
3 & \{\sigma_{002},\sigma_{120}\} \\ \hline
5 & \{\sigma_{020},\sigma_{102}\} \\ \hline
6 & \{\sigma_{022},\sigma_{100}\} \\ \hline
7 & \{\sigma_{200},\sigma_{322}\} \\ \hline
4 & \{\sigma_{202},\sigma_{320}\} \\ \hline
2 & \{\sigma_{220},\sigma_{302}\} \\ \hline
1 & \{\sigma_{222},\sigma_{300}\} \\ \hline
 \hline
\end{array}$$
\end{table}

In table~\ref{tb:QBC2}  a {\em Quantum Multidirectional Communication Protocol} is sketched~\cite{Gao08,Chong11}. The purpose of this protocol is to communicate securely four classical bits, two emitted by Alice, one by Bob and another by Claire. By repeating the protocol the parties may exchange longer bit-strings. Alice, Bob and Claire should interchange four classical bits, two emitted by Alice, one by Bob and another by Claire.
The parties share two GHZ states, ${\bf c}_0={\bf c}_1={\bf b}_{\ell}\in\hache_3$ with respective components ${\bf c}_{00},{\bf c}_{10},{\bf c}_{20}$ and ${\bf c}_{01},{\bf c}_{11},{\bf c}_{21}$. The components ${\bf c}_{0k}$, ${\bf c}_{1k}$, ${\bf c}_{2k}$ are in possession of Alice, Bob and Claire respectively, $k\in\intv{0}{1}$. The quregister ${\bf c}_0$ is a record of the initial state ${\bf b}_{\ell}$, while the quregister ${\bf c}_1$ is to be transformed during the protocol.

\begin{table}
\caption{{\em Quantum Multidirectional Communication Protocol} \label{tb:QBC2}}
\centering\fbox{\begin{minipage}{.95\textwidth}
\begin{enumerate}
\item The two bits of Alice determine an index $i_A\in\intv{0}{3}$. She applies $\sigma_{i_A}$ to her qubit ${\bf c}_{01}$.
\item Bob applies either $\sigma_0$ or $\sigma_2$ to his qubit ${\bf c}_{11}$ according to the value of his bit.
\item Claire applies either $\sigma_0$ or $\sigma_2$ to her qubit ${\bf c}_{21}$ according to the value of her bit.
\item They take a measure of the transformed quregister with respect to the Bell basis.
\item Using table~\ref{tb:SA},  since each participant knows his/her own message, they recover the transmitted bits.
\end{enumerate}
\end{minipage}}
\end{table}

Another bidirectional protocol~\cite{Man06} consists of three participants: Alice and Bob are the correspondents and Claire is the controller. The correspondents are able to communicate only after the authorization of the controller, but their correspondence should be kept in secret against the controller. The protocol is sketched at table~\ref{tb:QCBCk}. Alice and Bob should interchange messages at $\intv{0}{3}^m$ after the authorization granted by Claire.
The parties share a constant sequence $\left({\bf c}_{\nu}\right)_{\nu=0}^{n-1}$ whose entries coincide with a GHZ initial state ${\bf b}_{\ell}$. The component sequence $\left({\bf c}_{0\nu}\right)_{\nu=0}^{n-1}$, let us say for ease of explanation, is owned by Claire, the component sequence $\left({\bf c}_{1\nu}\right)_{\nu=0}^{n-1}$ by Alice and the component sequence $\left({\bf c}_{2\nu}\right)_{\nu=0}^{n-1}$ by Bob.
\begin{table}
\caption{{\em Quantum Controlled Bidirectional Communication Protocol} \label{tb:QCBCk}}
\centering\fbox{\begin{minipage}{.95\textwidth}
\begin{enumerate}
\item Alice and Bob agree a set $J\subset\intv{0}{n-1}$ of $m$ positions among the index set $\intv{0}{m-1}$.
\item Alice codifies her message $\left(a_{\mu}\right)_{\mu=0}^{m-1}\in\intv{0}{3}^{m-1}$ by applying $\sigma_{a_{\mu}}$ to her correspondent qubit ${\bf c}_{1\nu_{\mu}}$, with $\nu_{\mu}\in J$, and she applies arbitrary Pauli operators at her qubits with indexes not in $J$. Alice sends her codified sequence to Claire.
\item Bob codifies his message $\left(b_{\mu}\right)_{\mu=0}^{m-1}\in\intv{0}{3}^{m-1}$ by applying $\sigma_{b_{\mu}}$ to his correspondent qubit ${\bf c}_{2\nu_{\mu}}$, with $\nu_{\mu}\in J$, and he applies arbitrary Pauli operators at his qubits with indexes not in $J$. Bob sends his codified sequence to Claire.
\item Claire receives the component sequences $\left({\bf c}_{1\nu_{\mu}}\right)_{\nu=0}^{n-1}$ and $\left({\bf c}_{2\nu_{\mu}}\right)_{\nu=0}^{n-1}$, and she measures the whole sequence $\left({\bf c}_{\nu_{\mu}}\right)_{\nu=0}^{n-1}$ with respect to the basis $B_3$. She sends her results to Alice and Bob as an authorization to proceed the transaction.
\item Using the table~\ref{tb:A3}, her knowledge of her own message and the index set $J$, Alice recovers Bob's message $\left(b_{\mu}\right)_{\mu=0}^{m-1}\in\intv{0}{3}^{m-1}$.
\item Using the table~\ref{tb:A3}, his knowledge of his own message and the index set $J$, Bob recovers Alice's message $\left(a_{\mu}\right)_{\mu=0}^{m-1}\in\intv{0}{3}^{m-1}$.
\end{enumerate}
\end{minipage}}
\end{table}

A {\em Key Agreement Protocol Using Entanglement Swapping} is obtained~\cite{Gao10} as follows. Let $B_2^{(\mu\nu)}$ be the Bell basis considering two qubits $\mu,\nu\in\intv{0}{3}$, $\mu\not=\nu$.

Let ${\bf c}^{(01)},{\bf c}^{(23)}$ be two Bell 2-quregisters with respective components ${\bf c}^{(0)},{\bf c}^{(1)}$ and ${\bf c}^{(2)},{\bf c}^{(3)}$.

Alice may act on the pair $({\bf c}^{(01)},{\bf c}^{(23)}) = [{\bf c}^{(0)},{\bf c}^{(1)},{\bf c}^{(2)},{\bf c}^{(3)}]$ either by ($A_0$:) doing nothing or by ($A_1$:) swapping the middle qubits, obtaining thus $[{\bf c}^{(0)},{\bf c}^{(2)},{\bf c}^{(1)},{\bf c}^{(3)}]$.

Bob may act on the pair $({\bf c}^{(01)},{\bf c}^{(23)}) = [{\bf c}^{(0)},{\bf c}^{(1)},{\bf c}^{(2)},{\bf c}^{(3)}]$ either by ($B_0$:) measuring $[{\bf c}^{(0)},{\bf c}^{(1)}]$ with respect to the Bell basis $B_2^{(01)}$ and measuring $[{\bf c}^{(1)},{\bf c}^{(2)}]$ with respect to the Bell basis $B_2^{(23)}$ or ($B_1$:) by measuring $[{\bf c}^{(0)},{\bf c}^{(2)}]$ with respect to the Bell basis $B_2^{(02)}$ and measuring $[{\bf c}^{(1)},{\bf c}^{(3)}]$ with respect to the Bell basis $B_2^{(13)}$.

If the chosen actions are $(A_0,B_0)$ or $(A_1,B_1)$, the actions are said to be {\em correlated}, otherwise, they are {\em anticorrelated}.

\begin{table}
\caption{{\em Key Agreement Protocol Using Entanglement Swapping} \label{tb:kapues}}
\centering\fbox{\begin{minipage}{.95\textwidth}
\begin{enumerate}
\item Alice selects a sequence ${\bf B} = \left({\bf b}_{\alpha(k)}\right)_{k=0}^{2m-1}$ of entangled states at the Bell basis. Each pair of two such states $\left({\bf b}_{\alpha(2k)},{\bf b}_{\alpha(2k+1)}\right)$, $k\in\intv{0}{m-1}$, involves 4 qubits, say $[{\bf c}^{(0k)},{\bf c}^{(1k)},{\bf c}^{(2k)},{\bf c}^{(3k)}]$
\item For each $k\in\intv{0}{m-1}$, Alice applies an operation $A_0$ or $A_1$ to the current pair $\left({\bf b}_{\alpha(2k)},{\bf b}_{\alpha(2k+1)}\right)$. She obtains the sequence ${\bf C}$ and she sends this sequence to Bob through a public quantum channel.
\item Bob receives the sequence ${\bf C}$ and for any block of 4 consecutive qubits, say $[{\bf d}^{(0k)},{\bf d}^{(1k)},{\bf d}^{(2k)},{\bf d}^{(3k)}]$, he selects randomly an operation $B_0$ or $B_1$ and he applies it to $[{\bf d}^{(0k)},{\bf d}^{(1k)},{\bf d}^{(2k)},{\bf d}^{(3k)}]$.
\item Alice and Bob exchange through a classic channel the lists of their corresponding selected operations.
\item The common key is established by selecting the 4 blocks measurements corresponding to the correlated pairs of operations. 
\item It is worth to remark that at the anticorrelated positions, both Alice and Bob may recover two common bits, corresponding to the initial state of Alice for the current 4-block.
\end{enumerate}
\end{minipage}}
\end{table}

In the protocol, the agreed common key is the juxtaposition of the measures obtained at the positions in which correlated operators do occur. When looking for a greater efficiency it is possible to recover also not 4, for 2 bits at any block corresponding to an anti correlated operator.

\end{document}